\documentclass[useAMS,usenatbib]{mn2e}
\usepackage{graphicx}
\usepackage{graphics,float}                         
\usepackage{epsf}
\usepackage{amsmath}                                
\usepackage{amssymb}
\usepackage{picinpar}                               
\usepackage{relsize,setspace,subfigure}
\usepackage{tabularx}
\usepackage[innercaption]{sidecap}                  
\usepackage[figuresright]{rotating}
\usepackage{url}
\usepackage{aas_macros}
\usepackage{natbib}
\include{aajournals}

\def\apjl{ApJ}%
\def\apjs{ApJS}%
\def\ao{Appl.~Opt.}%
\def\aap{A\& A}%
\onecolumn

\title[An Alternative Way To Model Presentday Universe]{Bianchi I Model: An Alternative Way To Model The Presentday Universe}
\author[Russell, Battal K{\i}l{\i}n\c{c} and Pashaev]{Esra Russell$^{1,2}$\thanks{E-mail: esrarussell@iyte.edu.tr}, Can Battal K{\i}l{\i}n\c{c}$^{3}$ and Oktay K. Pashaev$^{1}$
\\
$^{1}$Izmir Institute of Technology, Department of Mathematics, 35430, Gulbahcekoyu/Urla, Izmir, Turkey.\\
$^{2}$Kapteyn Astronomical Institute, P.O. Box 800, 9700 AV Groningen the Netherlands.\\
$^{3}$Ege University, Department of Astronomy and Space Sciences, 35100, Bornova, Izmir, Turkey.}
\begin{document}
\date{Accepted .... Received ...; in original form ...}

\pagerange{\pageref{firstpage}--\pageref{lastpage}} \pubyear{2013}

\maketitle

\label{firstpage}

\begin{abstract}
Although the new era of high precision cosmology of the cosmic microwave background (CMB) radiation improves our knowledge to understand the infant as well as the presentday Universe, it also leads us to question the main assumption of the exact isotropy of the CMB. There are two pieces of observational evidence demonstrating that there is no exact isotropy. These are first the existence of small anisotropy deviations from isotropy of the CMB radiation and second, the presence of large angle anomalies that are shown as real features by the Planck satellite results. These evidences are particularly important since isotropy is one of the two main postulates of the Copernican principle on which the FRW models are built. This almost isotropic CMB radiation implies that the universe is almost a FRW universe, as is proved by previous studies.

Assuming the matter component forms the deviations from isotropy in the CMB density fluctuations when matter and radiation decouples, we here attempt to find possible constraints on these deviations by using the Bianchi type I (BI) anisotropic model which is asymptotically equivalent to the standard FRW. Then we obtain the separations from the FRW type scale factor and the Hubble parameter in the form of time dependent functions. Hence we put constraints on the anisotropy coefficients from the anisotropy upper limits of the recent Planck data and from the theoretical consistency relation. These constraints lead us to obtain a BI model which becomes to be an almost FRW in time that is consistent with observational data of the CMB.
\end{abstract}


\section{Introduction}\label{intro}
The accepted model of the presentday Universe is homogenous and isotropic on large scales and is defined by the Robertson Walker (RW) metric in the de Sitter space time. On the other hand, the new high precision cosmology era brings some new insights to our understanding of the infant universe via two important pieces of observational evidences of broken isotropy of the cosmic microwave background (CMB) radiation. The first evidence for broken isotropy is small temperature anisotropies with approximately $10^{-5}$ amplitude. The second evidence is the family of some large angle anomalies \citep{2003ApJS..148...97B} that are the alignment of quadrupole and octopole moments
\citep{2005PhRvD..72j1302L,2004IJMPD..13.1857R,2004PhRvD..69f3516D,2004PhRvD..70d3515C}, large scale asymmetry
\citep{2004ApJ...605...14E,2004MNRAS.354..641H}, the unusually cold spot
\citep{2004ApJ...609...22V} and the low quadrupole moment. Although there are
various studies to explain the cosmological origin of the large angle anomalies such as an\-i\-so\-trop\-ic inflation \citep{2004JCAP...10..016B,2005PhRvD..72j3002G,2007JCAP...09..006P,2007JCAP...11..005E,2008JCAP...08..005Y,2008JCAP...08..021K},
inhomogeneous spaces \citep{2006ApJ...644..701J,2006MNRAS.367.1714L,2007MNRAS.377.1473B,2007MNRAS.380.1387P},
local spherical voids \citep{2006ApJ...648...23I}, an initial phase of inflation
\citep{2003JCAP...07..002C,2009PhRvD..80b3526D} and a non-trivial spherical topology
\citep{2003Natur.425..593L}, it was not certain that the anomalies were observational artefacts. However, recent data from the Planck satellite prove that these anomalies represent real features of the CMB map of the Universe \citep{planck2013}. This result has a key importance since the small temperature anisotropies and large angle anomalies may be caused by some unknown mechanisms or an anisotropic phase during the evolution of the Universe. This statement is particularly interesting since it aids us to modify the current model or to construct an alternative model to decode the effects of the early universe on the presentday large scale structure without affecting the processes in the nucleosythesis.

The answer of how to construct a new universe model or modify the standard model may be found in the early theories. According to the theories proposed by \cite{1968ApJ...151..431M} and \cite{1977PhRvD..15.2738G}, anisotropy of the early stage of the universe may turn into an isotropic present universe and initial anisotropies die away. As is known from the presentday observational data there are anisotropies in the CMB, therefore it is possible that the anisotropies or anomalies are imprints of this early anisotropic phase on the CMB. Apart from this, \cite{Stoeger1995} and \cite{MaartensEllisStoeger1996} show if all fundamental observers measure the CMB radiation to be almost isotropic in an expanding universe, the universe is locally almost spatially homogeneous and isotropic in a region. This result formalizes the way of an almost FRW spacetime since the time of the decoupling of matter and radiation based on the evidence almost-isotropy of the CMB \citep{HawkingEllis,Stoeger1995,MaartensEllisStoeger1996,Cea2014arXiv1401.5627C}. One may question the affects of an almost FRW model on the primordial nucleosynthesis.  \cite{Barrow1997PhRvD..55.7451B} shows that it is possible for anisotropic fluids to create a measurable temperature anisotropy in the CMB without having any significant effects upon the primordial nucleosynthesis of $He^{4}$. In fact, \cite{CampanelliII} discuss this matter and confirm Barrow's result.

Bianchi models can be alternatives to the standard FRW models with small deviations from the exact isotropy in order to explain the
anisotropies and anomalies in the CMB. In this frame work,
\cite{2005ApJ...629L...1J} propose that the large angle anomalies can be mimicked by using a specific solution of the
Bianchi type VII$_{h}$ universe based on models developed by \cite{1973MNRAS.162..307C} and
\cite{1985Natur.316...48B}. In addition, \cite{2007MNRAS.380.1387P} point out that the polarization signal in the Bianchi type VII$_{h}$ universe can mimic the several large angle anomalous features observed in the CMB. However, \cite{2013arXiv1303.5086P} show that the Bianchi type VII$_{h}$ model is not consistent with the observational data from the Planck satellite. Apart from this, one of the large angle anomalies of the CMB; the low quadrupole moment is particularly important since it may indicate a presence of Bianchi type I (BI) anisotropic evolution in the early universe. How this happens? The low quadrupole moment shows a great amount of power suppression at large scales. This suppression cannot be satisfied by the standard dark energy dominated cold dark matter model as in indicated by the recent Planck and the WMAP results \citep{2013arXiv1303.5075P}. Particularly \cite{2013arXiv1303.5075P} underline that the standard model is incomplete. Early, \cite{MartinezSanz95} prove that the small quadrupole component of the CMB temperature found by COBE implies that if the universe is homogeneous but anisotropic BI then it necessarily must be a small departure from the flat Friedmann model. Later on, \cite{Cea2014arXiv1401.5627C,CampanelliII} and \cite{CampanelliI} show that the low quadrupole moment can be reduced in a plane symmetric BI universe. Recently, \cite{2013JCAP...12..003A} analyze the state space of a BI universe with anisotropic sources. In their study, assuming the universe contains anisotropic model including matter and dark energy components since decoupling, they find that this type of BI model contributes dominantly to the CMB quadrupole. Given its importance for studying the possible effects of an anisotropy in the early universe on presentday observations, many researchers have investigated the BI model from different perspectives. Examples of these studies are, string theory \citep{2001PhLB..509..151A,2008Ap&SS.317...79R,2009Ap&SS.321...37R,2010IJMPA..25.3043B,2012Ap&SS.339..371R},
dynamical properties
\citep{1983NCimB..77...62S,MonicaF,2006GReGr..38.1003E,Akarsu2010,Akarsu2010GReGr,2011IJTP...50.2720A,2012IJTP...51...90S,ericlinder2012,2013Ap&SS.343..423M,2013Ap&SS.343..489P,2013IJTP...52...69S,2013IJTP...52.4055A,2013PhRvD..88f3518K},
the singularity problem \citep{bronnikovchudayevashikin, Khalatnikov, Belinskij}, the spinor/scalar field \citep{Saha,2001PhRvD..64l3501S,2005GReGr..37.1233F,SB,2005Ap&SS.299..149S,saha06,2012GReGr..44.1893K,2012IJTP...51.3769P,2013arXiv1302.1354S,2013CQGra..30t5010C}
and perturbations in the early phase of inflation \citep{2007JCAP...11..005E,
2007JCAP...09..006P,2010AcASn..51..109D,2011IJTP...50.3043B,2012MPLA...2750014A,2011JCAP...05..011K}. Separately, \cite{MaartensEllisStoegera1995} obtain evolution equations by which matter imposes anisotropies on freely propagating CMB radiation, leading to a new model independent way of using anisotropy measurements to limit the deviations of the Universe from a FRW geometry. Following this, \cite{MaartensEllisStoeger1996} show how to place the quadrupole and octopole direct and explicit limits on the shear, vorticity, Weyl tensor and density gradients. Later, \cite{SAGnE} point out that it is possible to find limits on all anisotropy parameters such as shear, viscosity, Weyl tensor and density gradients,...etc., if one can determine limits on the anisotropy components from the CMB measurements.

In this study, our main goals here are to construct an anisotropic BI model that leads us to obtain deviations from the standard FRW one and is to limit the most up to date constraints via the two distortion/shear parameters which are directly obtained from the limits on the dipole, quadrupole and octopole of the recent Planck data. Then, assuming that the CMB anisotropies show themselves as slight deviations from isotropy in the density fluctuations since decoupling of matter and radiation, we formulate possible constraints on the separations from isotropy by modeling the evolution of the BI model that is asymptotically the FRW one. Then we obtain deviations in the form of time dependent functions. As a result we put constraints on these deviations by using the data on the anisotropy upper limits from the Planck satellite and the consistency relation that is provided by previous studies.

The structure of this paper is the following. First, we give the general framework of the BI model
including the necessary theoretical background and the isotropization criteria of the BI model to the FRW one at decoupling. After providing the well known exact solutions of the field equations of the BI model, the criteria of deviations with two different dynamical behaviors are obtained which are directly related with the sign of deviation/anisotropy parameters. Then we calculate the upper limits of the deviations satisfying the average and late time distortion/shear values obtained from the Planck anisotropy limits. Finally the discussion and conclusions are summarized.

\section{BI models}\label{sec:1}
The Bianchi type I model admits the metric element that has different scale factors in each direction,

\begin{eqnarray}
dl^2={c^2}dt^2-a^{2}_{1}(t)dx^2-a^{2}_{2}(t)dy^2-a^{2}_{3}(t)dz^2,
\label{generalmetric2}
\end{eqnarray}
\noindent
where $a_{1}$, $a_{2}$ and $a_{3}$ represent three different scale factors which are a function of time $t$. If we admit the energy-momentum tensor of a perfect fluid, then the field equations of the BI universe are found as,

\begin{subequations}
\label{feforgm}
\begin{eqnarray}
\frac{{\dot{a}_{1}}{\dot{a}_{2}}}{a_{1} a_{2}}+\frac{{\dot{a}_{1}}{\dot{a}_{3}}}{a_{1}
a_{3}}+\frac{{\dot{a}_{2}}{\dot{a}_{3}}}{a_{2} a_{3}}&=&{8\pi G}\rho,\\
\frac{{\ddot{a}_{1}}}{a_{1}}+\frac{{\ddot{a}_{3}}}{a_{3}}+\frac{{\dot{a}_{1}}{\dot{a}_{3}}}{a_{1} a_{3}}&=&
-\frac{8\pi G}{c^2}p,\\
\frac{{\ddot{a}_{2}}}{a_{2}}+\frac{{\ddot{a}_{1}}}{a_{1}}+\frac{{\dot{a}_{2}}{\dot{a}_{1}}}{a_{2}
a_{1}}&=&-\frac{8\pi G}{c^2}p,\\
\frac{{\ddot{a}_{3}}}{a_{3}}+\frac{{\ddot{a}_{2}}}{a_{2}}+\frac{{\dot{a}_{3}}{\dot{a}_{2}}}{a_{3}
a_{2}}&=&-\frac{8\pi G}{c^2}p.
\end{eqnarray}
\end{subequations}
\noindent
Here the dot represents the derivatives in terms of time $t$. To solve the system of equations (\ref{feforgm}), we define the following new variables which are simply the
directional Hubble parameters,

\begin{eqnarray}
{H_{1}}\equiv\frac{{\dot{a}_{1}}}{a_{1}},
\phantom{a}
{H_{2}}\equiv\frac{{\dot{a}_{2}}}{a_{3}},
\phantom{a}
H_{3}\equiv\frac{{\dot{a}_{3}}}{a_{3}}.
\label{transformationgenera1}
\end{eqnarray}
\noindent
Their first derivatives are,

\begin{eqnarray}
\dot{H}_{1}=\frac{{\ddot{a}_{1}}}{a_{1}}-\left(\frac{{\dot{a}_{1}}}{a_{1}}\right)^2,
\phantom{a}
\dot{H}_{2}=\frac{{\ddot{a}_{2}}}{a_{2}}-\left(\frac{{\dot{a}_{2}}}{a_{2}}\right)^2,
\phantom{a}
\dot{H}_{3}=\frac{{\ddot{a}_{3}}}{a_{3}}-\left(\frac{{\dot{a}_{3}}}{a_{3}}\right)^2.
\label{transformationgenera2}
\end{eqnarray}
\noindent
Inserting variables (\ref{transformationgenera1}) and their derivatives (\ref{transformationgenera2}) into the
Einstein field equations (\ref{feforgm}), we reformulate the field equations in terms of the directional Hubble parameters,

\begin{subequations}
\label{frgm11}
\begin{eqnarray}
{H}_{1} {H}_{2} +{H}_{1}H_{3}+{H}_{2} H_{3} &=& {8\pi G}\rho,\label{constraintH}\\
\dot{H}_{3}+H^2_{3}+\dot{H}_{1}+{H}^2_{1}+H_{3} {H}_{1} &=&-\frac{8\pi G}{c^2}p,\\
\dot{{H}}_{1}+{H}^2_{1}+\dot{H}_{2}+{H}^2_{2}+{H}_{1} {H}_{2}&=&-\frac{8\pi G}{c^2}p,\\
\dot{H}_{3}+{H}^2_{3}+\dot{H}_{2}+{H}^2_{2}+{H}_{3} {H}_{2} &=&-\frac{8\pi G}{c^2}p.
\end{eqnarray}
\end{subequations}
\noindent
In addition to this, the energy conservation equation $T^{\mu}_{\nu;\mu}=0$ yields,
\begin{eqnarray}
\dot{\rho}&=&-\left(\frac{{\dot{a}_{1}}}{a_{1}}+\frac{{\dot{a}_{2}}}{a_{2}}+\frac{{\dot{a}_{3}}}{a_{3}}\right)
\left(\rho+\frac{p}{c^2}\right)=-3H\left(\rho+\frac{p}{c^2}\right).
\label{vaneinsteintensor}
\end{eqnarray}
\noindent
As is known, the BI universe has a flat metric with $k=0$ which implies that its total density is equal
to the critical density. The critical density is given by,

\begin{eqnarray}
\label{criticalBIden}
\rho=\rho_{c} &=&\frac{1}{8\pi G}\left({H}_{1} {H}_{2} +{H}_{1}H_{3}+{H}_{2} H_{3}\right).
\end{eqnarray}

\subsection{General Solution}
In this subsection, we derive the analytical solutions of the field equations of the BI models in terms of
the directional Hubble parameters. To do this, first we add the last three equations of system (\ref{frgm11}),
which yields,
\small
\begin{eqnarray}
\label{gensola}2\frac{d}{dt}\left(\sum_{i=1}^{3}
H_{i}\right)+2\left(H^2_{1}+H^2_{2}+H^2_{3}\right)+
\left({H}_{3} {H}_{2}+{H}_{1}{H}_{2}+{H}_{3} {H}_{1}\right)=\frac{-24\pi G}{c^2}p.
\end{eqnarray}
\normalsize
After substituting the following term,

\begin{eqnarray}
\sum_{i=1}^{3} H^{2}_{i}=\left(\sum_{i=1}^{3} H_{i}\right)^{2}-2\left({H}_{3}
{H}_{2}+{H}_{1}{H}_{2}+{H}_{3}
{H}_{1}\right),
\end{eqnarray}
\noindent
and equation (\ref{constraintH}) of system (\ref{frgm11}) into equation (\ref{gensola}), we then obtain,

\begin{eqnarray}
\frac{d}{dt}\left(\sum_{i=1}^{3} H_{i}\right)+{\left(\sum_{i=1}^{3} H_{i}\right)^2}={12\pi
G}\left(\rho-\frac{p}{c^2}\right).
\label{rvs}
\end{eqnarray}
\noindent
The mean of the three directional Hubble parameters in the BI universe is given by,

\begin{eqnarray}
H\equiv\frac{1}{3}\left({H}_{1}+{H}_{2}+{H}_{3}\right)=\frac{1}{3}\left(\frac{{\dot{a}_{1}}}{a_{1}}+
\frac{{\dot{a}_{2}}}{a_{2}}+\frac{{\dot{a}_{3}}}{a_{3}}\right).
\label{meanratefirst}
\end{eqnarray}
\noindent
Substituting the mean (\ref{meanratefirst}) into equation (\ref{rvs}), a nonlinear first order
differential equation is obtained,

\begin{eqnarray}
\dot{H}+3H^2={4\pi G}\left(\rho-\frac{p}{c^2}\right).
\label{riccati}
\end{eqnarray}
\noindent
Here this dynamical equation shows evolution of the Hubble parameter of the related BI cosmology.

In addition to this, it is possible to write equation (\ref{riccati}) in terms of volume element $V$ by using
the following relation between volume and the mean Hubble parameter of the BI,

\begin{eqnarray}
H=\frac{1}{3}\frac{d}{dt}\ln(a_{1} a_{2} a_{3})=\frac{1}{3}\frac{\dot{V}}{V}.
\label{evvolume}
\end{eqnarray}
\noindent
As is seen, the multiplication of the scale factors in different directions is defined as the volume element of
the BI universe $V\equiv a_{1} a_{2} a_{3}$. Using this relation between volume and the mean Hubble parameter in equation (\ref{riccati}), the volume evolution equation of the BI models is obtained,

\begin{eqnarray}
\ddot{V}-3\left[4\pi G (\rho-\frac{p}{c^2})\right] V=0.
\label{riccativolume}
\end{eqnarray}
\noindent
On the basis of the above, we find the following alternative form for system (\ref{frgm11}),

\begin{subequations}
\label{generalsolutionofanisotropic}
\begin{eqnarray}
\dot{H}_{1}+3{H}_{1}H={4\pi G}\left(\rho-\frac{p}{c^2}\right),\label{generalsolutionofanisotropic1}\\
\dot{H}_{2}+3{H}_{2} H={4\pi G}\left(\rho-\frac{p}{c^2}\right),\label{generalsolutionofanisotropic3}\\
\dot{H}_{3}+3{H}_{3} H={4\pi G}\left(\rho-\frac{p}{c^2}\right).\label{generalsolutionofanisotropic2}
\end{eqnarray}
\end{subequations}
\noindent
These expressions allow us to write down the generic solution of the directional Hubble parameters,

\begin{eqnarray}
{H}_{i}(t)&=&\frac{1}{\mu(t)}\left[K_{i}+ \int \mu(t){4\pi
G}\left(\rho(t)-\frac{p(t)}{c^2}\right)dt\right],\phantom{a}i=1, 2, 3,
\label{Hubbleparameters}
\end{eqnarray}
\noindent
where $K_{i}$s are the integration constants. The integration factor $\mu$ is defined as,

\begin{eqnarray}
\mu(t)=e^{\int^{t} 3H(s)ds}.
\label{integfact}
\end{eqnarray}
\noindent
The integration factor $\mu$ in the solutions (\ref{Hubbleparameters}) is derived from the system
(\ref{generalsolutionofanisotropic}) by the particular solution of the system itself.

As can be seen from solutions (\ref{generalsolutionofanisotropic}), the initial values/integration constants
determine the solution of each directional Hubble parameter. These values are the origin of the anisotropy. Note
that the generic solution of the directional Hubble parameters (\ref{Hubbleparameters}) is incomplete. To obtain
exact solutions of the Hubble parameters and therefore the Einstein equations, we need one more equation which
is known as the equation of state,

\begin{eqnarray}p=\gamma \rho c^2.
\label{eos}
\end{eqnarray}
Here the adiabatic parameter $\gamma$ is characterized by a component of the universe dominating its expansion,

\begin{description}
           \item[.] $\gamma= 1/3$, radiation dominated universe,
           \item[.] $\gamma=0$, matter dominated universe.
         \end{description}
Before giving the analytical formalisms of the different epochs of the BI models, it is useful to define a
general isotropy criterion of BI type Universe models that is essential to obtain asymptotically FRW ones.

\section{Isotropization of BI Models into FRW Universe}\label{Sec:Isotropization}
The isotropic and homogeneous nature of the large scale structure may be an asymptotic
situation emerging from an an\-i\-so\-trop\-ic nature of the universe is formed by the matter component during decoupling. That is why it is essential to define an isotropizaton criteria which should explain how the anisotropy parameters disappear or become negligible when the Universe evolves into the present epoch, $t\rightarrow t_{0}$.

\cite{bronnikovchudayevashikin} and \cite{saha06,saha09} define isotropization as expansion factors of the BI
universe that grow at the same rate at late stages of the evolution. It is assumed that a BI model becomes
isotropic if the ratio of each directional expansion factor $a_{i}(t)$ ($i=1$, $2$, $3$) and the expansion
factor of the total volume $a(t)$ tends to be a constant value,

\begin{eqnarray}
\frac{a_{i}}{a}\rightarrow \text{constant} > 0\phantom{a}\text{when}\phantom{a} t\rightarrow \infty.
\label{conditionisot}
\end{eqnarray}
\noindent
Note that the total expansion factor $a(t)$ has the contribution from each directional expansion factor,

\begin{eqnarray}
a=\left(a_{1}a_{2}a_{3}\right)^{1/3}=V^{1/3}.
\end{eqnarray}
\noindent
The an\-i\-so\-trop\-ic models satisfying condition (\ref{conditionisot}) become isotropic. As a particular case of condition (\ref{conditionisot}), one may choose the constant as unity. This indicates that when the universe evolves into the presentday $t=t_{0}$, its dynamics become equivalent to the FRW. As a consequence of this choice, even highly anisotropic the BI models become isotropic in time. Since our goal is to obtain a model that becomes a FRW one in the late time limit, then it should be specifically indicated that the an\-i\-so\-trop\-ic parts of each scale factor of the BI models should tend to
be identical and equivalent to unity in the limit of presentday $t=t_{0}$. Under this condition, the critical
density of the BI models is reduced to,

\begin{eqnarray}
\frac{\rho}{\rho_{c}} &=&\Omega=1,\phantom{a} \rho_{c}= \frac{3{H_{0}^2}}{8\pi G},
\label{criticaldenBI}
\end{eqnarray}
\noindent
where the total mean density $\Omega$ of the Bianchi models is unity due to the flat geometry of the BI metric
(\ref{generalmetric2}). Apart from this general isotropization criteria (\ref{conditionisot}), we consider the
following two widely used anisotropy criteria in the literature
\citep{jacobs,bronnikovchudayevashikin,saha06,saha09},

\begin{eqnarray}
A&=&\frac{1}{3}\sum_{i=1}^{3}\left(\frac{H^2_{i}-{H^2}}{{H^2}}\right)\rightarrow 0, \label{anistpara}\\
\sigma^2&=&\frac{3}{2}A H^2\rightarrow 0,
\label{shear}
\end{eqnarray}
\noindent
where $A$ is the mean anisotropy parameter while $\sigma^2$ is the shear scalar and it is defined as,

\begin{eqnarray}
\sigma^2\equiv\sigma_{ij}\sigma^{ij},
\label{shaertensortraceless}
\end{eqnarray}
\noindent
where $\sigma_{ij}$ is the shear tensor. The shear tensor indicates any tendency of distortion  into an ellipsoidal shape of the initially spherical region. Therefore, the shear scalar $\sigma^2$ represents the distortion rate of the region. The mean anisotropy parameter $A$ in equation (\ref{anistpara}) is correlated with the expansion divergence $\Theta$ also known as expansion scalar which is related to the expansion rate/Hubble parameter as,

\begin{eqnarray}
\Theta= \nabla. \vec{v}=3H,
\end{eqnarray}
\noindent
this leads to,

\begin{eqnarray}
\Theta\equiv \frac{1}{3}\frac{\nabla. \vec{v}}{3H}=\sum_{i=1}^{3}\left(\frac{H_{i}-1}{H}\right),
\end{eqnarray}
\noindent
in which $v$ represents the velocity field. Note that the isotropy of every point of the Universe implies that the
vorticity $\omega$ and shear $\sigma$ of the matter are zero \citep{1973MNRAS.162..307C}. The vorticity $\omega$
is the rate of rotation of a set of axes fixed in the matter, relative to a set of inertial axes defined by
gyroscopes. In the BI universe the vorticity parameter is zero since there is no rotation by definition. The
distortion rate is the difference between the Hubble parameters of the matter in each
orthogonal direction, and its value is nonzero for the BI models.
These parameters, $A$ and $\sigma^2$, have two constraints; one of them is theoretical and it directly
comes from the consistency relation of the analytical solution of the field equations of the BI metric
(\ref{feforgm}) via the integration constants
\citep{jacobs,bronnikovchudayevashikin,saha06,saha09,2007JCAP...09..006P,GuptSingh2012PhRvD..86b4034G,Amirhashchi2014Ap&SS.tmp..103A,Saha2014IJTP...53.1109S},

\begin{eqnarray}
\sum_{i=1}^{3} K_{i}=0.
\label{Kis}
\end{eqnarray}
\noindent
The second constraint is the observational value of the shear parameter (\ref{shear}).
\cite{1996PhRvL..77.2883B} analyzed four year data of the Cosmic Background Explorer (COBE) satellite to
constraint the allowed parameters of the Bianchi type VII$_{h}$ model. This model evolves into a FRW Universe and the definitive upper limits on the amount of shear, $\left[\sigma/H\right]_{0}$ and vorticity, $\left[\omega/H\right]_{0}$ are,

\begin{eqnarray}
\left[\frac{\sigma}{H}\right]_{0} & < & 3\times 10^{-9},\\
\left[\frac{\omega}{H}\right]_{0} & < & 10^{-6},\phantom{a} \Omega_{0} = 1.
\end{eqnarray}
\noindent
Apart from the above limits, \cite{Stoeger1995} show how to relate the CMB anisotropies to growing density in homogeneties in an almost-FRW expanding universe. Later on, \cite{MaartensEllisStoegera1995,MaartensEllisStoegerb} relate CMB anisotropies with anisotropies and inhomogeneities in the large scale structure of the universe and show the way of placing limits on those anisotropies and inhomogeneities simply by using CMB quadrupole and octopole limits. Note that these limits are upper bounds on the multipoles of the CMB temperature anisotropy and are independent from any models for the source of perturbations including inflationary models \citep{MaartensEllisStoegerb,MaartensEllisStoegera1995,1999ApJ...522..559S}. Hence, the distortion/shear is limited as follows \citep{MaartensEllisStoegera1995,MaartensEllisStoegerb},

\begin{eqnarray}
\frac{|\sigma_{ij}|}{\Theta} < \frac{5}{3}\epsilon_{1} +  {3}\epsilon_{2} + \frac{3}{7} \epsilon_{3},
\label{stoegerlimitonshear}
\end{eqnarray}
\noindent
in which $\epsilon$'s stands for limits on the dipole, the quadrupole and the octopole components respectively. Therefore, if one can obtain $\epsilon$'s in equation (\ref{stoegerlimitonshear}) from the CMB measurements then one can determine the upper bond of the distortion/shear parameter by relating the squares of the rotationally invariant rms multipole $\Delta T^{2}_{l}$ coefficients in the usual Legendre polynomial expansion as,

\begin{eqnarray}
\langle \epsilon_{l}\rangle^2=3^{l}\frac{\left(2l\right)!}{2^{l}\left(l !\right)^2} \Delta T^{2}_{l},
\label{Gebbie}
\end{eqnarray}
\noindent
in which $\Delta T^2_{l}$ are directly related with the anisotropy rms such as $\Delta T^2_{2}=Q^{2}_{rms}$ and  $\Delta T^2_{3}=O^{2}_{rms}$ in which $Q_{rms}$ and $O_{rms}$ are the rms quadrupole and octopole amplitudes that are obtained from the CMB observations as is shown by \cite{1994ApJ...436..423B} that $\left(\Delta T_{l}\right)^2$ are compatible with the COBE data. Using (\ref{Gebbie}), the average quadrupole $ \langle\epsilon_{2}\rangle $ and octopole $\langle\epsilon_{3}\rangle$ limits are found in terms of rms, which are,

\begin{eqnarray}
\langle \epsilon_{2}\rangle=\sqrt{13.5}\frac{Q_{rms}}{T_{0}},
\label{epsilon1a}
\end{eqnarray}
\noindent
\begin{eqnarray}
\langle \epsilon_{3}\rangle=\sqrt{67.5}\frac{O_{rms}}{T_{0}},
\label{epsilon1b}
\end{eqnarray}
\noindent
where $T_{0}=2.7255$K is the average temperature of the CMB since \cite{SAGnE,1999ApJ...522..559S} obtain the
multipoles for $\Delta T/T$, ${Q_{rms}}=10.7\pm 7$ $\mu$K \cite{1994ApJ...436..423B,1996ApJ...464L...5K} and
$O_{rms}=16\pm 8$ $\mu$K are the best-fit value, rms quadrupole and octopole amplitudes of COBE in which the
dipole rms is set as zero $\epsilon_{1}=0$ \citep{SAGnE,1999ApJ...522..559S}. Then,
\cite{SAGnE,1999ApJ...522..559S} find average distortion by using calculated anisotropy limits in
(\ref{epsilon1a}) and (\ref{epsilon1b}) in the shear equation (\ref{stoegerlimitonshear}) based on the the COBE
data,

\begin{eqnarray}
\langle\frac{|\sigma_{ij}|}{\Theta}\rangle < 4.4\times10^{-5}.
\label{stoegerlimitonshear2}
\end{eqnarray}
\noindent
Apart from the distortion equation (\ref{stoegerlimitonshear}) which is a model
independent parameter, \cite{MaartensEllisStoegera1995} show that if the dipole, quadrupole and octopole limits
are homogeneous to first order, then distortion can be found directly without any further assumption,

\begin{eqnarray}
\left(\frac{|\sigma_{ij}|}{\Theta}\right)_{0}=\left(\frac{16}{15}\frac{\Omega_{r}}{\Omega_{m}}\right)_{0}\epsilon_{2}.
\label{distortion2}
\end{eqnarray}
\noindent
which is also defined as the late-time limit of the shear. As is seen above, the distortion only depends on quadrupole upper bound $\epsilon_{2}$ rather than octopole anisotropy limit. In distortion equation (\ref{distortion2}) $\Omega_{r}$ and $\Omega_{m}$ are the radiation and matter density parameters. Using this distortion equation (\ref{distortion2}), \cite{MaartensEllisStoeger1996} obtain the distortion to characterize the deviations from isotropy of a space time which has nearly BI symmetry for the COBE data in which the quadrupole is $\epsilon\approx 10^{-5}$ and $\left(\Omega_{r}/\Omega_{m}\right)_{0}=2.5 h^{-2}\times10^{-5}$ \citep{KolbTurner1990}, that is,

\begin{eqnarray}
\left(\frac{\sigma_{ij}}{\Theta}\right)_{0}= 2.7 h^{-2}\times 10^{-10},
\label{distortion2a}
\end{eqnarray}
\noindent
where the parameter $h$ is estimated as $0.4 < h < 1.0$.
%
\section{Evolution of Anisotropic Deviations from FRW in Decoupling}
In this section, we show that the given BI model provides a solution of expansion factors and Hubble parameters of the FRW model with extra parameters that decrease in time. These extra terms are defined as deviations from isotropy of the FRW model. Here we obtain functional forms of deviations from the FRW isotropic model by using anisotropic and homogeneous BI metric in order to investigate the dynamical characteristics of separations from isotropy during the decoupling. These deviations from isotropy  are possibly detected as tiny anisotropy fluctuations in the CMB resulting in observed anisotropy and inhomogeneities such as distortion even though anisotropies are very small.

As is expected, radiation and matter decouples during decoupling, that happens right after radiation- matter equality at $z_{eq}\approx 3300$ ($t\approx 10^{4}$ years) with a temperature of approximately $3\times 10^{5}$K. After this equality the temperature of the universe drops to $3000$K, and the plasma turns into neutral gas around $z< 3300$ ($t\approx 10^{4}$ years).

Since we here investigate deviations from isotropy by using the BI metric, our first starting point is to explain the dynamical behavior of the BI model in which matter and radiation components are decoupled, and later on, matter starts to rule the evolution. In such a period, the energy conservation equation has contributions from both matter and radiation components in the BI universe as the FRW one. As a result, the radiation and matter state equation decouples, in which the pressure term of the matter component vanishes due to its adiabatic parameter $\gamma=0$ while the radiation pressure becomes proportional to one third of the radiation
density in equation of state (\ref{eos}),

\begin{eqnarray}\label{eosrm}
{p}_{m}=0,\phantom{a}{p}_{r}=\frac{1}{3}{\rho}_{r}.
\end{eqnarray}
\noindent
Hence, the energy conservation equation (\ref{vaneinsteintensor}) in the radiation-matter period decouples
as well,

\begin{eqnarray}
\dot{\rho}_{r}=-4 H_{rm} {\rho}_{r},\phantom{a} \dot{\rho}_{m}=-3 H_{rm}{\rho}_{m},
\label{energyconsermatradzero}
\end{eqnarray}
\noindent
which leads to,
\begin{eqnarray}
{\rho}_{r}=\rho_{r,0}\frac{V_{rm,0}}{V_{rm}},\phantom{a} {\rho}_{m}=\rho_{m,0}\frac{V_{rm,0}}{V_{rm}}.
\label{energyconsermatrad}
\end{eqnarray}
\noindent
Here the radiation ${\rho}_{r}$ and matter ${\rho}_{m}$ densities as well as the volume element in the decoupling era $V_{rm}$ are normalized to their presentday values, $\rho_{r,0}$, $\rho_{m,0}$, $V_{rm,0}$ in which the presentday values of the volume element are equal to unity. The normalization of the densities (\ref{energyconsermatrad}) will help us to compare the dynamical parameters to the recent observational results. Then the normalized densities for two components are written by using the definition of the critical density (\ref{criticaldenBI}), which are,

\begin{eqnarray}
{\rho}_{r,0}=\rho_{c,0}\Omega_{r,0}=\frac{3H^2_{0}}{8\pi G}\Omega_{r,0}\phantom{a}\text{and}\phantom{a}{\rho}_{m,0}=\rho_{c,0}\Omega_{m,0}=\frac{3H^2_{0}}{8\pi G}\Omega_{m,0}.
\label{energyconsermatrad2}
\end{eqnarray}
\noindent
Our goal is to show deviations from isotropy by dynamical parameters such as Hubble parameter and expansion factor by using the BI model which has the FRW model embedded. Therefore, first the form of the mean Hubble parameter $H_{rm}$ which has the contributions from three directional Hubble parameters is obtained by using the dynamical evolution equation (\ref{riccati}) and equation of state (\ref{eosrm}) in order to obtain the exact solution of the directional Hubble parameters in the radiation $+$ matter dominated BI model from equations (\ref{Hubbleparameters}),

\begin{eqnarray}
\dot{H}_{rm}+3H^2_{rm}={4\pi G}\left(\frac{2}{3}\rho_{r}+\rho_{m}\right).
\label{riccatirm}
\end{eqnarray}
This nonlinear equation gives the dynamical evolution of the mean Hubble parameter at decoupling. Equation (\ref{riccatirm}) can be transformed into an equation that satisfies the total
volume evolution of the model by substituting relation (\ref{evvolume}) and the normalized radiation and matter densities (\ref{energyconsermatrad}), which is,

\begin{eqnarray}
{\ddot{V}_{rm}}-{4\pi G}\left(\rho_{m,0}+\frac{2}{3}\frac{\rho_{r,0}}{V_{rm}^{1/3}}\right)=0.
\label{mixturevolmatrad}
\end{eqnarray}
\noindent
Multiplying this equation with $\dot{V}_{rm}$, integrating it in terms of time, and substituting the normalized densities (\ref{energyconsermatrad2}) in, we then obtain,

\begin{eqnarray}
{\dot{V}}^2-9H^2_{0}\Omega_{m,0} V - 9H^2_{0}\Omega_{r,0} V^{2/3}=0.
\label{integrationofmixturerm}
\end{eqnarray}
\noindent
Here, we obtain a relation between the mean Hubble parameter and the volume element of the related epoch. This equation is rearranged as follows,
\begin{eqnarray}
\left(\frac{\dot{V}_{rm}}{V_{rm}}\right)^2=9 {H^2_{0}}\left(\frac{\Omega_{m,0}}{V_{rm}}+\frac{\Omega_{r,0}}{V_{rm}^{4/3}}\right).
\label{integrationofmixturematrad}
\end{eqnarray}
\noindent
The integration of the above equation allows us to derive a relation for the time component as a function
of volume during decoupling,

\begin{eqnarray}
{H_{0}}t=\frac{4}{3}\frac{V_{rme}^{2/3}}{\sqrt{1-\Omega_{m,0}}}
\left(\sqrt{1+\left(\frac{V_{rm}}{V_{rme}}\right)^{1/3}}
\left(\frac{1}{2}\left(\frac{V_{rm}}{V_{rme}}\right)^{1/3}-1\right)\right),
\label{htmixturematrad}
\end{eqnarray}
\noindent
where ${V_{rme}}\equiv\left(\frac{\Omega_{r,0}}{\Omega_{m,0}}\right)^{3}$. Hence, the volume element can be found,

\begin{eqnarray}
V_{rm}\approx \frac{9}{4} {H^2_{0}}{\Omega_{m,0}} t^{2} + 5\left(\frac{\Omega_{rad,0}}{\Omega_{m,0}}\right)^3= {V_{m}} + 5 {V_{rme}}.
\label{volumeradmatter}
\end{eqnarray}
\noindent
Therefore, the mean Hubble parameter is obtained as,

\begin{eqnarray}
H_{rm}\approx\frac{2}{3}\frac{1}{t}\frac{1}{\left[1+5\frac{V_{rme}}{V_{m}}\right]},
\label{mixturehubble}\end{eqnarray}
\noindent
by using relation (\ref{evvolume}). Here, in the limit of $V_{rm} \gg {V_{rme}} $ ($t\rightarrow\infty$),
the solution of equation (\ref{integrationofmixturematrad}) approaches the mean Hubble parameter of the matter
dominated BI universe,

\begin{eqnarray}
H_{rm}\rightarrow H_{m}=\frac{2}{3}\frac{1}{t}.
\end{eqnarray}
\noindent
Taking into account that the radiation and matter become equivalent when $V^{1/3}_{rme} = 2.963\times 10^{-4}$ \citep{2003itc..book.....R}, the redshift $z_{eq}$ when decoupling starts, can be obtained,

\begin{eqnarray}
\label{mradequitime}
\frac{\Omega_{m}}{\Omega_{r}}={V^{-1/3}_{rme}}\frac{a}{a_{0}}={\frac{\Omega_{m,0}}{\Omega_{r,0}}}\frac{1}{1+z},\phantom{a}
1+z_{eq}\approx 3.375\times 10^{3}.
\end{eqnarray}
\noindent
This result leads to the redshift value $z_{eq}\approx 3300$. In the FRW Universe, radiation matter equality
took place at a scale factor $a_{rm}\equiv \Omega_{r,0}/\Omega_{m,0}\approx 2.8\times 10^{-4}$. Here, we assume that the particles that are nonrelativistic today were also nonrelativistic
at $z_{eq}$; this should be a safe assumption, with the possible exception of massive neutrinos, which make a
minority contribution to the total density \citep{tc}. It follows that the integration factor of the epoch in
terms of volume element (\ref{volumeradmatter}) from equation (\ref{integfact}), is obtained,

\begin{eqnarray}
\mu_{rm} = 4 \Omega_{m,0}^3 V_{rm}.
\label{integfactorrm}
\end{eqnarray}
\noindent
Substituting the integration factor (\ref{integfactorrm}) and using the equation of state (\ref{eos}) for the decoupling case in the solution of the directional Hubble parameters (\ref{Hubbleparameters}), we obtain,

\begin{eqnarray}
H_{i}t_{0}=\underbrace{\frac{\alpha_{i}}{4 \Omega_{m,0}^3}\left(\frac{t_{0}}{t}\right)^2 b}_{\text{Deviations}} + \underbrace{\frac{2}{3}\left(\frac{t_{0}}{t}\right)b}_{\text{FRW:}\phantom{a}\text{Hubble}\phantom{a}\text{Parameter}},
\label{timesmrad}
\end{eqnarray}
\noindent
where the parameter $b$ and the normalized deviation/anisotropy coefficients $\alpha_{i}$ are defined as,

\begin{eqnarray}
b \equiv \frac{V_{m}}{V_{rm}},\phantom{a}\alpha_{i}\equiv \frac{{K_{i}}}{{t_{0}}}.
\end{eqnarray}
\noindent
As is seen in the directional Hubble parameters ($H_{1}$, $H_{2}$, $H_{3}$, inequations (\ref{timesmrad})), the last term on the right hand side is the same in each directional Hubble parameter which is the standard Hubble parameter of the FRW model at decoupling. The first terms on left hand side with the anisotropy coefficients $\alpha_{i}$ are only dependent on the initial values ${K_{i}}$ (see equations \ref{Hubbleparameters}). As is seen, even small differences in the expansion rates of the given epoch may cause deviations from the standard model. The normalized scale factors are derived from the directional Hubble parameters  (\ref{timesmrad}) with a
direct integration in terms of cosmic time in transformation (\ref{transformationgenera1}), which are obtained as,

\begin{eqnarray}
{a}_{n,i}
= \underbrace{e^{\left[\frac{\alpha_{i}}{4\sqrt{5} \beta} \tan^{-1}
\left({\beta}\frac{\left(1-\frac{t_{0}}{t}\right)}{1+5 \frac{V_{rme}}{V_{m}} \frac{t}{t_{0}}
}\right)\right]}}_{\delta_{n,i}:\phantom{a}\text{Deviations}}
\underbrace{\left[1+\left(\frac{\left(\frac{t}{t_{0}}\right)^2-1}{1+5
\frac{V_{rme}}{V_{m}}}\right)\right]^{1/3}}_{\text{FRW}},
\label{radmatterscalefactors2}
\end{eqnarray}
\noindent
where the density dependent parameter $\beta$ is defined as,
\begin{eqnarray}
\beta \equiv \left(\Omega_{r,0}\Omega_{m,0}\right)^{3/2}.
\end{eqnarray}
\noindent
Note that anisotropic parts of the scale factors or deviations from isotropy $\delta_{n,i}$ should satisfy, \citep{Saha2014IJTP...53.1109S,Amirhashchi2014Ap&SS.tmp..103A}

\begin{eqnarray}
\prod^{3}_{i=1} \delta_{n,i}=1,
\end{eqnarray}
\noindent
It is crucial to note that even though the BI model shows deviations from isotropic FRW in each direction, the overall volume of the universe behaves as the standard FRW model. As a result, the total volume element is not affected by the directional  deviations (expansion/contraction(s)) in the decoupling. One easily can prove that via the multiplication of scale factors (\ref{radmatterscalefactors2}), which is related to volume via definition (\ref{conditionisot}) and taking into account the consistency relation of integration constants (\ref{Kis}), the sum of the normalized anisotropy coefficients $\alpha_{i}$ disappears,

\begin{eqnarray}
K_{1}+K_{2}+K_{3}=0 \Longrightarrow \alpha_{1}+\alpha_{2}+\alpha_{3} = 0.
\label{criteriaK}
\end{eqnarray}
\noindent
As a result, the total volume element is not affected by the directional deviations in the decoupling. Therefore the total volume becomes the volume of the universe given by the FRW in the related epoch. Apart from this, the critical anisotropy coefficients are obtained from the first derivative test of the normalized scale factors
(\ref{radmatterscalefactors2}), at which points the directional Hubble parameters become zero. These critical coefficients are given by,

\begin{eqnarray}
\alpha_{i}= -\frac{8}{3}\Omega^3_{m,0}\left(\frac{t}{t_{0}}\right),
\label{criticalcoeff}
\end{eqnarray}
\noindent
in which $\Omega_{m,0}=0.3175$ \citep{2013arXiv1303.5076P}. Assuming deviations from isotropy starts in decoupling, the critical anisotropy value is calculated as $-4.5$ $10^{-7}$ at $t/t_{0}=5.2$
$10^{-6}$ which is approximately the radiation- matter equality $z_{eq}=3300$. Then we investigate the
dynamical behaviors of the directional normalized scale factors around these critical points. As a result, we
obtain expansion and contraction criteria of the scale factors in terms of the critical anisotropy
coefficients as follows,

\begin{subequations}
\label{criteriaofnegradmatter}
\begin{eqnarray}
\text{Expansion}\phantom{a}\alpha_{i}&=&
\begin{cases} \text{If}\phantom{a} \alpha_{i} < 0,\phantom{a} \text{then}\phantom{a} |\alpha_{i}| <
\phantom{a} \frac{8}{3}\Omega^3_{m,0}\left(\frac{t}{t_{0}}\right)\label{criteriaofnegradmatter1},
\\
\text{If}\phantom{a} \alpha_{i} >\phantom{a}0,\phantom{a}\text{then}\phantom{a} \alpha_{i} > \phantom{a}
-\frac{8}{3}\Omega^3_{m,0}\left(\frac{t}{t_{0}}\right).
\end{cases}
\\
\text{Contraction}\phantom{a}\alpha_{i}&=&
\begin{cases} \text{If}\phantom{a} \alpha_{i} < 0,\phantom{a}\text{then}\phantom{a}
|\alpha_{i}|>\phantom{a} \frac{8}{3}\Omega^3_{m,0}
\left(\frac{t}{t_{0}}\right).\label{criteriaofnegradmatter2}
\end{cases}
\end{eqnarray}
\end{subequations}
\noindent
These criteria of the critical values are obtained by applying the second derivative test to the scale factors (\ref{radmatterscalefactors2}). Therefore the criteria (\ref{criteriaofnegradmatter}) show the dynamical characteristics of the scale factors via a large set of anisotropy coefficients admitting positive and negative numbers. The physical interpretation of negative anisotropy coefficients is equivalent to contraction of the related scale factor(s) while positive anisotropy coefficients demonstrate expansion characteristics of the scale factor(s).

\begin{figure}
\centering
\vspace{80mm}
\hspace{60mm}
\includegraphics[width=0.45\textwidth]{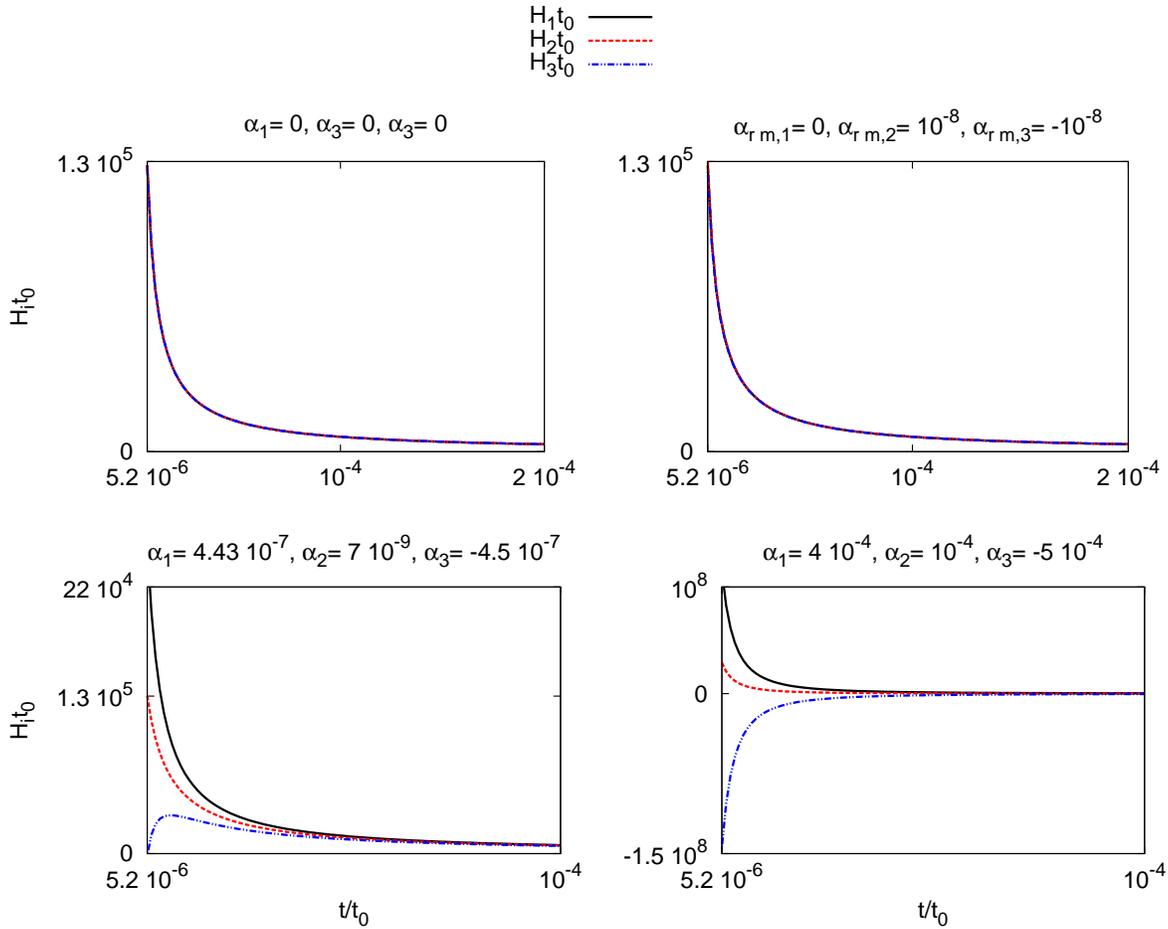}
\caption{Emerging of the negative anisotropy part $H_{3} < 0$ of the directional Hubble parameters
in the decoupling for the matter density $\Omega_{m,0}=0.3175$.}
\label{fig:mradhubs}
\end{figure}

\begin{figure}
\centering
\vspace{80mm}
\hspace{-40mm}
\includegraphics[width=0.6\textwidth]{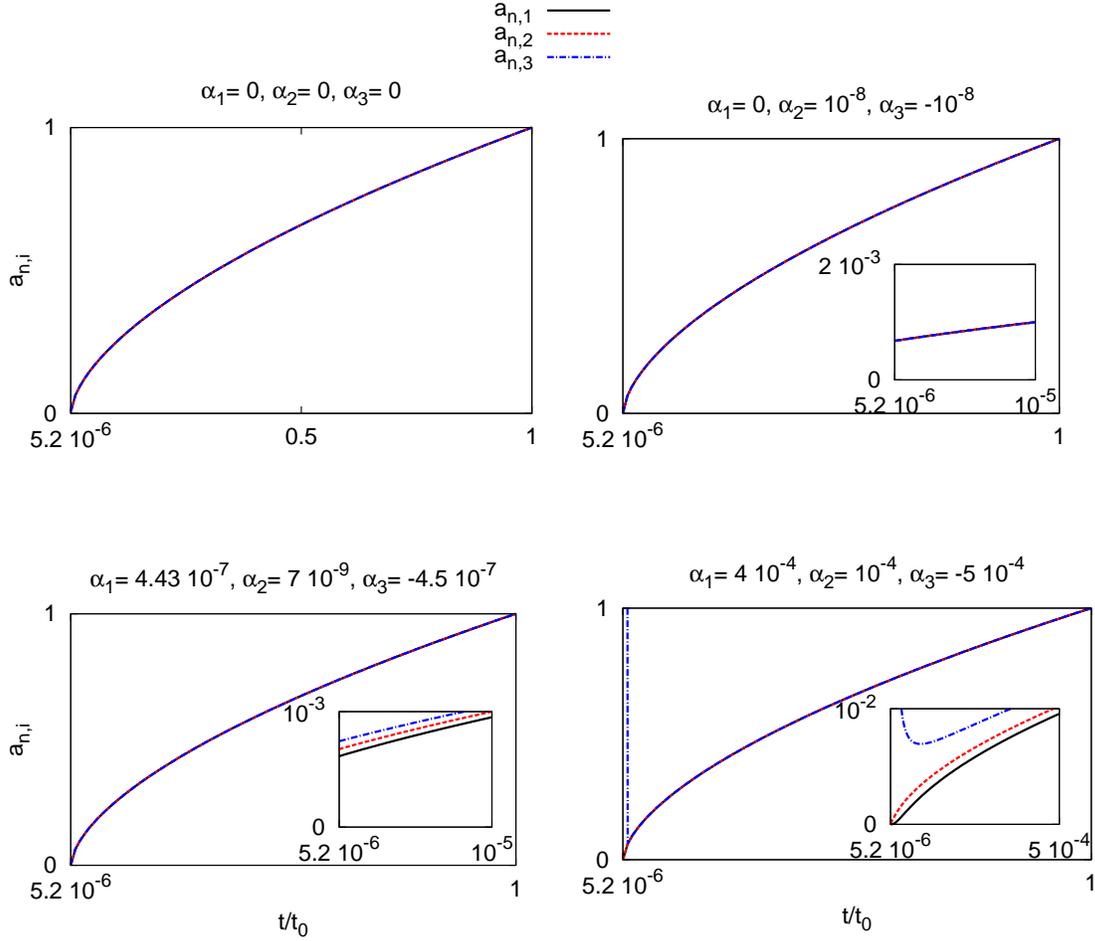}
\caption{The time evolution of the normalized scale factors starting from the decoupling with different sets of anisotropy coefficients $\alpha_{i}$.}
\label{fig:mrads}
\end{figure}

These initial expansions and/or contractions of the scale factors also affect the dynamical behaviors of
the directional Hubble parameters due to their dynamical relation. A possible contraction of one of the
directional scale factors causes slowing down in the rate of expansion leading to zero or even negative
expansion rates depending on how strong slowing down is in the related directions during the emerging of
anisotropies. On the other hand, anisotropic expansions of the directional scale factors cause increase in the
expansion rates. Therefore, criteria (\ref{criteriaofnegradmatter}) indicate another criteria on the directional Hubble parameters by presenting emerging out of positive and negative branches (adopting \cite{2007JCAP...11..005E}) of the expansion rates which are the same as expansion criterion (\ref{criteriaofnegradmatter1}) and contraction criterion (\ref{criteriaofnegradmatter2}) of the directional scale factors. Although anisotropies may form negative and/or positive branches of the directional Hubble parameters (or expansion and/or contraction of the directional scale factors) at the very early stages of the evolution, later on these negative/positive directional Hubble parameters (or contracting/expanding directional scale factors) tend to merge and become the Hubble parameter (the scale factor) of the FRW model in time see (Figs. \ref{fig:mradhubs}-\ref{fig:mrads})

Fig. \ref{fig:mradhubs} and Fig. \ref{fig:mrads} show how the directional Hubble and scale parameters change in
terms of time $t/t_{0}$ around and at the critical anisotropy coefficient relatively for the same set of coefficients. In Fig.
\ref{fig:mradhubs} the left upper panel with zero anisotropy coefficients presents the evolution of standard FRW Hubble parameter. Therefore the rest of the panels indicate separations from the evolution of the standard FRW Hubble parameter with non zero anisotropy coefficients. Particularly we aim to show the emerging of negative Hubble parameters in Fig. \ref{fig:mradhubs}. To show this evolution, the anisotropy coefficients of the two axes of expansion are kept positive ($\alpha_{1} > 0$ and $\alpha_{2} > 0$) from the upper right to the lower right panels. Note that positive anisotropy coefficients ($\alpha_{1} > 0$ and $\alpha_{2} > 0$) represent positive separations from the standard FRW Hubble parameter indicating initial speeding up process in the related directions at decoupling. On the other hand, in each panel (from upper right to lower right) the anisotropy coefficients of the third axis $H_{3}$ accept negative values (note that the sum of a set coefficients are zero satisfying the consistency relation \ref{Kis}). These negative valued directional Hubble parameters indicate different dynamical behaviors depending on comparison of the coefficients with the critical value based on the criteria (\ref{criteriaofnegradmatter}). The critical value is obtained as $-4.5\times 10^{-7}$ by using equation (\ref{criticalcoeff}) at initial time $t/t_{0}=5.2\times 10^{-6}$. Therefore, the third directional Hubble parameter ($H_{3}$) with the anisotropy coefficient that is equal to the exact critical value $-4.5\times 10^{-7}$ shows a halt in expansion or zero expansion rate at the beginning of decoupling $t/t_{0}=5.2\times 10^{-6}$ (see lower left panel). As is seen in the lower right panel the initially negative third directional Hubble parameter emerges since the anisotropy coefficients accepts $|\alpha_{i}| > 4.5\times10^{-7}$. Note that the negativity of directional Hubble parameters (contraction rate) gets stronger for the values of anisotropy coefficient higher than the critical anisotropy coefficient (see criterion \ref{criteriaofnegradmatter2}). Here after we call the positive and negative separations from standard FRW Hubble parameter as positive  and negative branches by following \cite{2007JCAP...11..005E}.

In Fig. \ref{fig:mrads}, the upper left panel shows the evolution of the standard FRW scale factor from decoupling to presentday. Panels from upper right to lower right demonstrate the evolution of the possible separations from the standard FRW. Here, the lower right panel shows a contraction until a certain time then this contraction turns into expansion in third direction. Note that directional scale factors especially with highly negative values satisfying the criteria (\ref{criteriaofnegradmatter2}) show initial contraction but, later on, this behavior starts turning into an expansion. This transformation of initial contraction to expansion is known as bouncing behavior of the directional scale factors in the BI models. When we choose any of the anisotropy coefficients with negative value $\alpha_{i} < 0$ satisfying the condition $|\alpha_{i}| > 4.5\times10^{-7}$, we can see this bouncing behavior of the related directional scale factor.

\subsection{Isotropization criteria of the radiation and matter dominated BI model}
\begin{figure}
\centering
\includegraphics[width=0.45\textwidth]{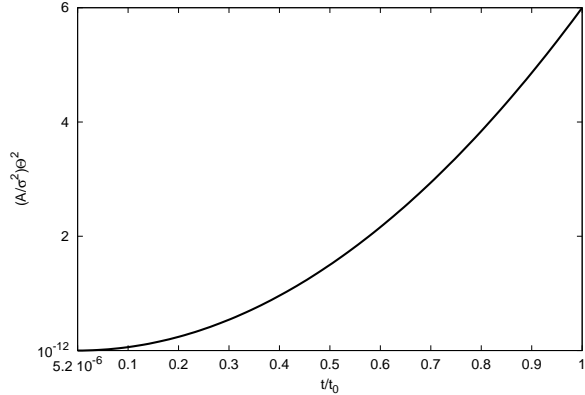}
\caption{Evolution of the ratio of the anisotropy-shear parameters starting from decoupling to the presentday $t=t_{0}$.}
\label{fig:aslm}
\end{figure}

Here, the isotropization criteria of the BI model at the decoupling are given by equations (\ref{anistpara}) and (\ref{shear}). Hence the isotropization of deviations from the exact isotropy is obtain by the anisotropy,

\begin{eqnarray}
\label{anisotrm}
A=\frac{3}{64 \Omega^6_{m,0}}{\left(\frac{t_{0}}{t}\right)^{2}}
{\sum^{3}_{i}\alpha_{i}^{2}}\rightarrow 0 & \mbox{if } t\rightarrow t_{0},
\end{eqnarray}
\noindent
and the shear parameters,
\begin{eqnarray}
\label{shearrm}
\left[\frac{\sigma}{\Theta}\right]^2=\frac{1}{128 \Omega^6_{m,0}}{\left(\frac{t_{0}}{t}\right)^{4}}
{\sum^{3}_{i}}\alpha_{i}^{2}\rightarrow 0 & \mbox{if } t\rightarrow t_{0}.
\end{eqnarray}
\noindent
Also, the ratio of anisotropy $A$ and the shear $\left[\frac{\sigma}{\Theta}\right]^2$ can be obtained for the same sum of squared anisotropy coefficients as,

\begin{eqnarray}
\label{shearradmratio}
\left(\frac{A}{\sigma^{2}}\right){\Theta}^2=6\left(\frac{t}{t_{0}}\right)^2.
\end{eqnarray}
\noindent
As is seen, at the presentday $t=t_{0}$, the ratio of anisotropy and distortion becomes $6$. This simple
formulation indicates that the expansion of the universe $\Theta$ increases during the evolution which is
consistent with observations \citep{Perlmutter,1998ApJ...507...46S,Riess98} while the ratio of pure anisotropy
parameters $\left(A/\sigma^2\right)$ should die away in time because of the expansion of the universe.
Considering the presentday Universe is isotropic on large scales, this is an expected result. Fig.
\ref{fig:aslm} shows the time evolution of the ratio of anisotropy and distortion parameters
(\ref{shearradmratio}) for the same sum of squared anisotropy coefficients. As is seen in Fig.
\ref{fig:aslm}, the anisotropy $\left(A/\sigma^2\right)$ is more dominant than expansion during the decoupling $t/t_{0}\approx 5.3\times 10^{-6}$ when matter creates the deviations
from isotropy. However, these
anisotropy parameters die away in time and the expansion of the universe takes over the evolution. Therefore,
the anisotropic BI behavior of the universe due to the matter component at decoupling turns into an isotropic
FRW one at the presentday.

\begin{figure}
\centering
\includegraphics[width=0.7\textwidth]{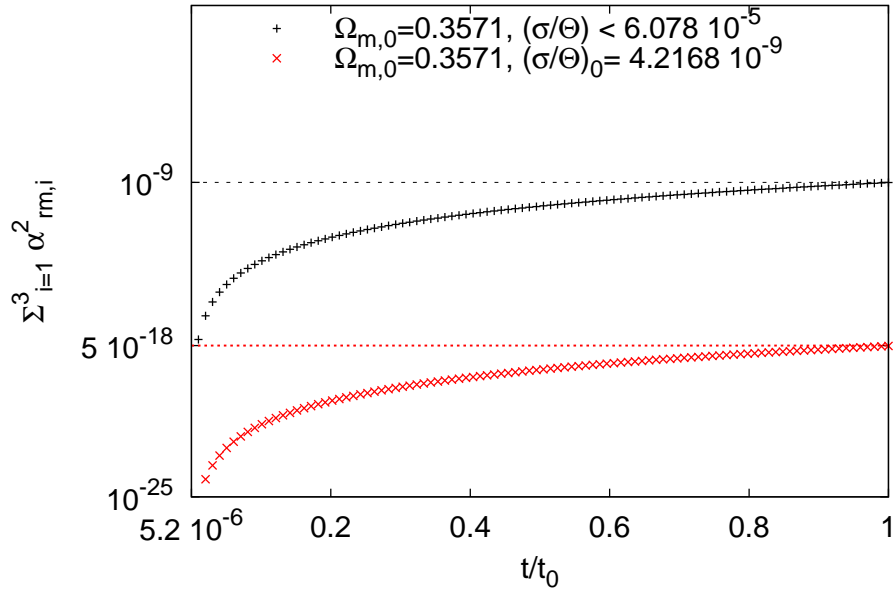}
\caption{Evolution of the sum of square anisotropy coefficients ${\sum^{3}_{i}}\alpha_{i}^{2}$ for the average distortion $6.078\times10^{-5}$ (black pluses) and the late time distortion $4.2168\times 10^{-9}$ (red crosses) as upper limits in the shear equation (\ref{shearrm}) starting from the decoupling.}
\label{fig:asrm}
\end{figure}

\noindent
Here we obtain an upper bound of the distortion based on anisotropy limits directly obtained from the recent Planck foreground-subtracted temperature power spectrum $\mathcal{D}_{l}$ \citep{2013arXiv1303.5076P} by adopting the same formalism of \cite{MaartensEllisStoegera1995,MaartensEllisStoegerb}. Note that the squares of rotationally invariant rms multipole moments $\left(\Delta T_{l}\right)^2$  should relate to each other in order to obtain correct multipole upper limits,

\begin{eqnarray}
\left({\Delta T}_{l}\right)^2= {\mathcal D}_{l}\frac{2l+1}{2l\left(l+1\right)},
\label{distortion2b}
\end{eqnarray}
\noindent
in which ${\mathcal D}_{l}$ is defined as,
\begin{eqnarray}
{\mathcal D}_{l}=\frac{l\left(l+1\right)}{2\pi} C_{l},\phantom{a}C_{l}= \frac{1}{2l+1}\sum^{+l}_{m=-l} |a_{lm}|^2,
\label{distortion2c}
\end{eqnarray}
\noindent
and $C_{l}$ is equivalent to the sum of the expansion coefficients $a_{lm}$ of the temperature anisotropy in spherical harmonics. The quadrupole $(l=2)$ and octopole $(l=3)$ anisotropy are given by \cite{2013arXiv1303.5076P} as,

\begin{eqnarray}
\label{Planckvalues}
{\mathcal D}_{2}\approx 299.5\phantom{a}[\mu K^2],\phantom{b} {\mathcal D}_{3}\approx 1000\phantom{a}[\mu K^2].
\end{eqnarray}
\noindent
According to this, the average quadrupole and octopole limits are found from the method of
\cite{MaartensEllisStoegerb,MaartensEllisStoegera1995} by using the quadrupole and octopole anisotropies of the
Planck data (\ref{Planckvalues}) in the average anisotropy definitions (\ref{epsilon1a}) and (\ref{epsilon1b}),
which are,

\begin{eqnarray}
\label{Planckvaluesquadandoctoplimit}
\langle{\epsilon}_{2}\rangle\approx 1.506\times 10^{-5},\phantom{a} \langle{\epsilon}_{3}\rangle\approx 3.640\times 10^{-5}.
\end{eqnarray}
\noindent
Substituting these values into distortion equation (\ref{stoegerlimitonshear}), we obtain distortion based on the Planck anisotropy from the power spectrum of the temperature fluctuations, which is,

\begin{eqnarray}
\label{Plancklimitdistortion}
\langle \frac{|\sigma_{ij}|}{\Theta}\rangle \phantom{a}< \phantom{a}6.078\times10^{-5}.
\end{eqnarray}
\noindent
Here we neglect the effect of the dipole component depending on \cite{Planck2014A&A...561A..97P} in which
a bulk flow has been significantly constrained by Planck studies of the kinetic Sunyaev-Zeldovich effect in which two different methods used to detect dipole as a consequence, in all cases the measured dipoles are compatible with zero. On the other hand, \cite{PlanckESM2013arXiv1303.5087P} report that our motion modulates and aberrates the CMB temperature fluctuations which is an order $10^{-3}$ effect applied to fluctuations which are already one part in roughly $10^{-5}$, so it is quite small. Nevertheless, it becomes detectable with the all-sky coverage, high angular resolution, and low noise levels of the Planck satellite. That is why we should be careful in order to construct an almost FRW model by using the CMB observed anisotropy limits. Apart from the average general shear limit (\ref{Plancklimitdistortion}), here we obtain the late-time limit of the shear which is also defined as the deviation from isotropy of a space time that is nearly BI symmetry based on the Planck data as,

\begin{eqnarray}
\label{Planckshearlatelimitvalue}
\left(\frac{\sigma_{ij}}{\Theta}\right)_{0} < 4.2168\times 10^{-9},
\end{eqnarray}
\noindent
where we choose $\left(\Omega_{r}/\Omega_{m}\right)_{0}=2.8 \times10^{-4}$ for the BI symmetry based on equation (\ref{mradequitime}).
Substituting the upper limits (\ref{Planckshearlatelimitvalue}-\ref{Plancklimitdistortion}) into the shear equation (\ref{shearrm}), one can find upper limits on the sum of squared anisotropy coefficients as a time dependent parameter. This leads us to obtain the two strong constraints on the deviations from isotropy. As a result, we can obtain a set of parameters at a given time and given observed shear parameter. In Fig. \ref{fig:asrm}, the upper limits of the sum of the square of the anisotropy coefficients $\sum_{i=1}^{3}\alpha^2_{rm,i}$ from equation (\ref{shearrm}) for the two upper distortion limits from Planck \citep{2013arXiv1303.5076P}. According to this, the sum of the square anisotropy coefficients indicates a very small limit value $5\times 10^{-18}$ at the presentday $t=t_{0}$ assuming the late time upper limit of the shear while the sum reaches $10^{-9}$ for the average distortion.

\section{Summary and Discussion}
Considering the observational evidence of the small anisotropic temperature fluctuations
and the presence of the large angle anomalies in the CMB, we here aim to construct emerging of the deviations from isotropy at the decoupling in which matter component is believed to form these separations following
\cite{HawkingEllis,Stoeger1995,MaartensEllisStoeger1996,Cea2014arXiv1401.5627C}. Also we attempt to put the
most stringent upper limits on these separations by using the recent Planck anisotropy upper bounds \cite{2013arXiv1303.5076P} taking into account the two shear formalisms that are derived from evolution equations in decoupling based on a model independent method proposed by \cite{MaartensEllisStoegera1995,MaartensEllisStoegerb}.

To construct the separations starting from the decoupling, first we consider anisotropic, homogeneous BI spacetime which is the most general case of the isotropic, homogeneous, flat FRW spacetime. Following this, we introduce the isotropization criteria by using the anisotropy (\ref{anistpara}) and shear (\ref{shear}) parameters from the previous studies \citep{jacobs,bronnikovchudayevashikin,saha06,saha09}. These criteria are particularly important since they tell us how the separations from isotropy emerge out of the BI spacetime, and later on, die out or become so small. We also introduce the two model independent formalisms of the shear/distortion \citep{MaartensEllisStoegera1995,MaartensEllisStoegerb}. These model independent shears are called the average shear and the late-time shear that are given in equations (\ref{stoegerlimitonshear}) and (\ref{distortion2}).

Apart from the isotropization criteria, we obtain solutions of the three directional Hubble parameters and the scale factors in the anisotropic spacetime, in which normalized deviations from the FRW type Hubble and scale factors are explicitly found as in equations (\ref{timesmrad}) and (\ref{radmatterscalefactors2}). Following this, the critical deviation/anisotropy coefficients (\ref{criticalcoeff}) are obtained in the time dependent functional forms. Depending on the sign of anisotropy coefficients $\alpha_{i}$ and their comparisons with the critical coefficient value at a given time, the directional scale factors can present contraction and/or expansion type separation(s) from the FRW scale factor. To generalize these two different dynamical characteristics of the directional scale factors, we formulate criteria of the expansion-contraction of the scale factors (\ref{criteriaofnegradmatter}).

According to criteria (\ref{criteriaofnegradmatter}), it is found that the scale factors with anisotropy coefficients satisfying expansion criterion (\ref{criteriaofnegradmatter1}) show expansion type separations from the FRW type isotropy during the early phase of decoupling. On the other hand, anisotropy coefficients satisfying criterion (\ref{criteriaofnegradmatter2}) demonstrate separations from the FRW scale factor with a contraction behavior in the related direction. However, these early expansion/contraction dynamical behavior tends to die away or become so small in time. As a result, any contraction and expansion type of separations in the scale factors may turn into FRW type expansion. This tendency is shown in Fig. \ref{fig:mrads}. Therefore, the scale factor of the anisotropic direction experiences a bounce due to initial contraction and soon later expansion in Fig. \ref{fig:mrads}. This behavior is also mentioned in \cite{2007JCAP...11..005E} and \cite{2008PhRvD..78j3525G} for a BI universe.

Moreover, criteria (\ref{criteriaofnegradmatter}) of the scale factors  are extended to the directional Hubble parameters in order to investigate their initial dynamical characteristic at decoupling. Then it is found that a possible contraction of one of the directional scale factors causes the rate of expansion to stop or slow down, which leads to zero or even negative expansion rates depending on the strength of the slowing down in the related directions. As a result, criteria (\ref{criteriaofnegradmatter}) turn into another criteria on emerging of initial positive and negative branches of the directional Hubble parameters (see in Fig. \ref{fig:mradhubs}). Therefore, the expansion criterion (\ref{criteriaofnegradmatter1})
becomes an indicator of emerging of a positive Hubble parameter while criterion (\ref{criteriaofnegradmatter2})
turns into an indicator of emerging of a negative Hubble parameter in the related direction at the given time. Here it is crucial to note that we do not have any observational evidence for the contracting scale factors or slowing down in expansion rates. Moreover, observational data support expansion of the universe. That is why we believe that one may exclude the contraction criterion, therefore the negative separation/anisotropy coefficients, in order to construct a model that is consistent with observations. On the other hand, as we pointed out before, the emerging of BI anisotropy tends to become an almost FRW isotropy in time. As a result, we should not rule out a possible scenario of slight initial contractions at the beginning of decoupling which later on, turn into the standard expansion characteristics of the FRW model.

Furthermore, the criteria (\ref{criteriaofnegradmatter}) with the consistency relation (\ref{Kis})
lead us to find set of coefficients form the initial conditions of the BI spacetime at decoupling that turns into an isotropic FRW model. Another constraint on the anisotropy coefficients comes from the observational data. Here we calculate the average shear as $6.078\times10^{-5}$ and the late time shear as $4.2168\times 10^{-9}$ by using the upper anisotropy limits for the quadrupole and octopole components from the recent Planck temperature power spectrum by following \cite{MaartensEllisStoegera1995,MaartensEllisStoegerb,Stoeger1995}. Then we use these average and the late time distortions in the distortion/shear equation (\ref{shearrm}) in order to obtain upper limits of the sum of the squared of the anisotropy coefficients for a given normalized time (or redshift) for the two distortions found from the Planck data. As a result, the values of the sum of square of anisotropy parameters calculated at the present time $t=t_{0}$ as $5\times 10^{-18}$ assuming the late time distortion upper limit while this value is $10^{-9}$ for the average distortion (Fig. \ref{fig:asrm}). These upper shear values indicate very small anisotropy coefficients that represent small deviations from the standard FRW. This result  is in agreement with \cite{MartinezSanz95}. They prove that if the universe is BI then it necessarily must be a small departure from the flat Friedmann model. Extending this result to the Planck satellite and point out that a possible BI dynamical behavior of the universe at decoupling leads to a small separations from isotropy of the FRW with the upper limits as $5\times 10^{-18}$ and $10^{-9}$ for the late-time shear and average shear respectively.

In short, here we obtain the model based on the most up to date constraints for the average and the late time shear parameters from the recent Planck data. These upper limits lead us to construct the deviations from isotropy in the dynamical form of the BI model turning into the FRW one in time. As a result, we construct the most stringent model to date that is consistent with recent CMB observations.


\bibliographystyle{mn2e}
\bibliography{bianchiImodels}
\bsp
\label{lastpage}
\end{document}